# The electronic structure of GaN and Ga investigated by soft x-ray spectroscopy and first-principles methods

Martin Magnuson[1], Maurizio Mattesini[2], Carina Höglund[1], Jens Birch[1] and Lars Hultman[1]

[1]*Department of Physics, Chemistry and Biology (IFM), Thin Film Physics Division, Linköping University, SE-58183 Linköping, Sweden*

[2]*Departamento de Física de la Tierra, Astronomía y Astrofísica I, Universidad Complutense de Madrid, Madrid, E-28040, Spain.*

**Abstract**

The electronic structure and chemical bonding of wurtzite-GaN investigated by N 1*s* soft x-ray absorption spectroscopy and N *K*, Ga $M_1$, and Ga $M_{2,3}$ emission spectroscopy is compared to that of pure Ga. The measurements are interpreted by calculated spectra using *first-principles* density-functional theory (DFT) including dipole transition matrix elements and additional on-site Coulomb interaction (WC-GGA+U). The Ga 4*p* - N 2*p* and Ga 4*s* - N 2*p* hybridization and chemical bond regions are identified at the top of the valence band between -1.0 and -2.0 and further down between -5.5 and -6.5 eV, respectively. In addition, N 2*s* - N 2*p* - Ga 4*s* and N 2*s* - N 2*p* - Ga 3*d* hybridization regions occur at the bottom of the valence band between -13 and -15 eV, and between -17.0 and -18.0 eV, respectively. A band-like satellite feature is also found around -10 eV in the Ga $M_1$ and Ga $M_{2,3}$ emission from GaN, but is absent in pure Ga and the calculated ground state spectra. The difference between the identified spectroscopic features of GaN and Ga are discussed in relation to the various hybridization regions calculated within band-structure methods.

# 1 Introduction

Among the group III nitrides, wurtzite gallium nitride (*w*-GaN) is an important wide band gap semiconductor compound for modern electronics. GaN has a relatively large direct band gap (3.4 eV), exhibit piezoelectricity, high thermal conductivity, mechanical strength and low electron affinity [1,2]. These properties constitute an excellent material platform for improved performance characteristics in optoelectronic device applications, e.g., LEDs in the UV-range, field emitters for flat displays, and high-speed transistors. In parallel, the scientific efforts are also concentrated on further understanding of the fundamental electronic structure properties of the nitrides and on improving of the crystalline quality by PVD and CVD methods. However, there is a lack of detailed understanding concerning the fundamental electronic structure and chemical bonding of wide bandgap nitride compounds. In *w*-GaN, there are two different types of Ga-N chemical bonds with π and σ character that have slightly different bond lengths.

Previous electronic structure measurements of GaN mainly include valence band x-ray photoelectron spectroscopy (XPS), which requires preparation of atomically clean and crystalline surfaces that are difficult to attain [3,4,5]. In XPS and UPS, the superimposed total contribution from the N 2*s*, N 2*p* and the Ga 4*s*, Ga 4*p* states of the valence band and the Ga 3*d* semicore states are probed. Soft x-ray absorption (SXA) and emission (SXE) spectroscopy employing dipole selection rules has also been applied at the N *K*-edge of GaN [6,7,8]. The top of the valence band in GaN is dominated by N 2*p* states whereas the Ga 4*s* states are much weaker and located at somewhat lower energy. SXE of GaN has been utilized at the Ga $L_3$ edge, but the position and shape of the weak Ga 4*s* valence states could not be resolved or identified [9]. Contrary to the case of the wide band gap nitride AlN [10], where the populated A 3*d* states (here A denotes the IIIB element Al) are located at the very top of the valence band, the Ga 3*d* states in GaN are located at the bottom of the valence band and therefore the N 2*p* - A 3*d* interaction plays a different role than in AlN. In N *K* emission of GaN, a small peak structure has been observed at the bottom of the valence band around -17 to -19 eV [9,11] and its intensity was found to depend on the Ga content [8]. However, the electronic structure and chemical bonding of N with Ga is still largely unknown in GaN. This includes, in particular, the Ga 4*p* states at the top of the valence band, the N 2*s* hybridization with the Ga 4*s* states at the bottom of the valence band, and the potentially important role of the shallow Ga 3*d* semicore states.

In this paper, we investigate in detail the bulk electronic structure and the influence of hybridization and chemical bonding between the N and Ga atoms of *w*-GaN in comparison to pure Ga. We apply angle-dependent SXA in the bulk-sensitive total fluorescence yield (TFY) mode and resonant and nonresonant element-specific SXE spectroscopy at the N 1*s*, Ga 3*s* and Ga 3*p* absorption edges to obtain information about the N 2*p*, Ga 4*s* and Ga 4*p* valence states, the role of the shallow Ga 3*d* semicore level and the hybridization with the N 2*s* states. We demonstrate that the wide band gap behavior of GaN is connected to specific changes in the electronic band structure, charge occupation and chemical bond regions in comparison to pure Ga. The experimental results are compared to electronic band structure calculations based on the generalized gradient approximation method of Wu and Cohen [12,13] (WC-GGA) including the optimized Coulomb interaction (U) which here accounts for the strong electronic interactions of the narrow Ga 3*d* band with localized electrons. The WC-GGA+U functional method is shown here to increase the calculated band gap and to improve considerably the computed Ga $M_{2,3}$ emission spectra as well as the bulk modulus of the wurtzite GaN phase. Moreover, the role of the N 2*p* - Ga 3*d* coupling for the magnitude of the band gap is discussed in comparison to other wide band gap nitrides such as AlN.

## 2 Experimental

### 2.1 Deposition and growth of GaN

A ~3.8 μm thick single-crystal stoichiometric *w*-GaN(0001) film was grown epitaxially by hot-wall metalorganic chemical vapor deposition (MOCVD) on a semi-insulating 4H-SiC substrate. A monocrystalline ~100 nm AlN nucleation layer was used to mitigate the lattice misfit between the SiC substrate and the GaN layer. The structure was grown while the substrate was rotating with a total reactor pressure of 50 mbar in a mixture of H and N carrier gases. Ammonia, NH, as a precursor for N, and trimethylgallium (TMGa) as precursor for Ga were supplied at a molar ratio of 385 with a NH flow-rate of 4 standard liters per minute and a susceptor temperature of 1000C [14,15].

The crystal structures of the GaN and pure Ga samples were characterized by X-ray diffraction (XRD) using Cu-K radiation. Figure 1 (top panel), shows XRD patterns of the GaN sample where the main scan is an overview Θ/2Θ scan (in Bragg-Brentano geometry) over a wide angular range. Due to the epitaxial nature of the material, only basal plane *w*-GaN 0002 and 0004 reflections are observed along with the SiC substrate 0004 and 0008 peaks and a single 0002 peak of the AlN seed layer. A high-resolution ω-scan of the GaN 0002 peak was performed with a Philips XPERT MRD diffractometer equipped with a symmetric 4xGe220 monochromator and a symmetric 3xGe220 triple-axis analyzer. The FWHM of the GaN 0002 reflection in the ω-direction is very narrow, only 60 arc-seconds showing that the GaN epilayer is of very high crystal quality. We determined the *w*-

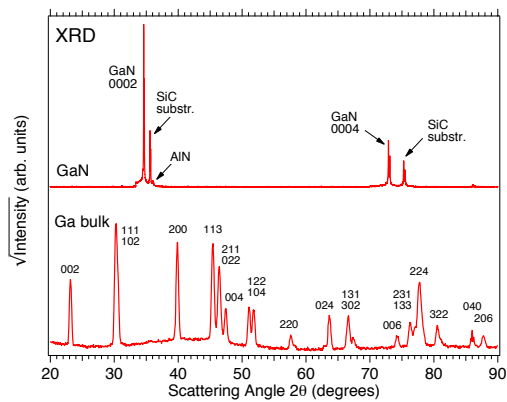

**Figure 1:** X-ray diffractograms (XRD) from the GaN (0001) thin film sample in comparison to pure Ga. The sharpness of the 0002 reflection is 60.3 arc-seconds from Rocking curve measurements which shows that the GaN coating is of very high epitaxial crystal quality.

GaN unit cell parameters to be $c$=5.1842 (±0.0009) Å and $a$=3.19207 (±0.00004) Å, by the method of Fewster and Andrews [16] using a high resolution asymmetric 4xGe220 monochromator and symmetric 3xGe220 analyzer setup.

XRD from pure Ga was performed in a Philips MRD system operated in a low-resolution parallel-beam configuration. The Ga (99.99999%) was cast into a flat platelet and filed clean with a diamond file just prior to the measurement during which it was kept at 28°C and flushed in dry N gas to prevent melting and limit oxidation. The overview scan in the lower panel of Fig 1 shows, indexed, all reflections expected from orthorombic (o) Ga [17]. The detected relative intensities are generally in very good agreement with all intensities in the reference powder data set with only minor deviations around the 200, 122, and 224 reflections, showing that the Ga sample is a randomly oriented polycrystalline material.

## 2.2 X-ray emission and absorption measurements

The SXE and SXA measurements were performed at the undulator beamline I511-3 at MAX II (MAX-lab National Laboratory, Lund University, Sweden), comprising a 49-pole undulator and a modified SX-700 plane grating monochromator [18]. The SXE spectra were measured with a high-resolution Rowland-mount grazing-incidence grating spectrometer [19] with a two-dimensional multichannel detector. The N $K$ SXE spectra were recorded using a spherical grating with 1200 lines/mm of 5 m radius in the first order of diffraction. The Ga $M_1$ and $M_{2,3}$ spectra were recorded using a grating with 300 lines/mm, of 3 m radius in the first order of diffraction. The SXA spectra at the N 1$s$ edges were measured both at normal incidence (90) and 20 grazing incidence with 0.08 eV resolution using the bulk-sensitive TFY mode (NEXAFS detector). During the N $K$, and Ga $M_1$, $M_{2,3}$ SXE measurements, the resolutions of the beamline monochromator were 0.2, 0.2 and 0.1 eV, respectively. The N $K$, Ga $M_1$ and Ga $M_{2,3}$ SXE spectra were recorded with spectrometer resolutions of 0.2, 0.3, and 0.06 eV, respectively. All measurements were performed with a base pressure lower than 5x10$^{-9}$ Torr. In order to investigate the anisotropy of the occupied N 2$p$ states, angle-dependent SXE measurements were made both at 20° and 70° incidence angles from the surface plane. The x-ray photons were detected parallel to the polarization vector of the incoming beam in order to minimize elastic scattering.

# 3 First-principles calculations

## 3.1 Electronic and structural properties

The SXA and SXE spectra were calculated with the wien2k code [20] employing the density-functional [21,22] augmented plane wave plus local orbital (APW+lo) computational scheme [23]. The Kohn-Sham equations were solved for the wurtzite (B4) GaN phase ($w$-GaN) using the WC-GGA functional for the exchange-correlation potential. Our electronic band structure calculations were carried out by employing the optimized WC-GGA $a$-lattice parameter and $c/a$ ratio. For the hexagonal close-packed lattice of *w*-GaN, we obtained *a*=3.1904 Å and *c*/*a*=1.6270, which are in excellent agreement with our experimental values (*a*=3.1921 and *c*/*a*=1.6241), previous experimental values (*a*=3.190 Å and *c*/*a*=1.627 [24,25]) and previous theoretical works [26]. The fitting of the energy-unit cell volume data by a Birch-Murnaghan [27] equation of state gave a bulk modulus of 193 GPa (B=4.464), that is also in reasonable agreement with the experimental bulk modulus of 210 GPa [28].

The WC-GGA functional works well for structural properties, but it has shown important limitations in reproducing the correct electronic band-structure of *w*-GaN. For instance, we compute a band-gap (E) of 1.83 eV, which is clearly underestimated with respect to the reported experimental value of 3.4 eV [24]. We have also found noticeable deficiencies in reproducing the correct energy position of the Ga 3*d* semicore states. This latter shortcoming is of major importance for this study as a non adequate description of the occupied 3*d*-states energy position implies a mismatch with the measured Ga $M_{2,3}$ edges. However, such a specific DFT-WC-GGA deficiency can be overcome by introducing a strong on-site Coulomb interaction in the localized Ga 3*d* bands. An effective Coulomb exchange interaction U=U-J was therefore used in these calculations, where U is the Coulomb energetic cost to place two electrons at the same site, and J is an approximation of the Stoner exchange integral. We treated the U value as a tunable parameter within the WC-GGA+U method and found that a value of 10 eV is needed to match up the theoretical Ga $M_{2,3}$ edges together with the experimental SXE spectra. Our optimized $U_{eff}$ parameter agrees fairly well with the U-value (11 eV) proposed by Cherian *et al*. [29] for the zinc-blende GaN structure. An *ad hoc* U-searching procedure was performed by first anchoring the computed Ga 4*s*-states to the energy positions of the same states in the experimental spectrum, and then varying the U-parameter up to the point when the theoretical Ga 3*d* doublet coincides with the measured one. The repositioning of the Ga 3*d* bands by about 4.0 eV downward in energy leads to a perfect agreement with the experimental binding energy of 17.5 ±0.1 eV [5]. The same energy movement was also detected at the bottom of the valence band for the N *K*-edge when adding the same Coulomb exchange interaction to the N 2*p* states, thus pointing to an important degree of hybridization between the Ga 3*d* and N 2*p* states. For the pure Ga system, we applied the same U-searching procedure as for GaN in order to correct for the wrong WC-GGA positioning of the 3*d* states and obtained an $U_{eff}$ of 6.0 eV. This smaller value can be understood in terms of the more delocalized (i.e. metallic) nature of pure gallium, where the DFT-WC-GGA method works better.

## 3.2 Theoretical SXA and SXE spectra

The computations of the SXE spectra were carried out within the so-called *final-state rule* [30], where no core-hole was created at the photo-excited atom. Theoretical emission spectra were computed at the converged WC-GGA+U ground-state density by multiplying the orbital projected partial density of states (pDOS) with the energy dependent dipole matrix-element between the core and the valence band states [31]. A comparison with the experimental spectra was achieved by including an instrumental broadening in the form of a Gaussian function and a final state lifetime broadening by convolution with an energy-dependent Lorentzian function with a broadening increasing linearly with the distance from the Fermi level.

The N *K* SXA spectra were calculated within the WC-GGA functional by including core-hole effects, thus specifically considering a crystal potential created from a static screening of the core-hole by the valence electrons. Such a self-consistent-field potential was generated in a 2×2×2 hexagonal supercell of 32 atoms (16 independent) containing one core-hole on the investigated element. The electron-neutrality of the system was kept constant by means of a negative background charge. By this procedure we explicitly include the excitonic coupling between the screened core-hole and the conduction electrons, which is an important requisite for simulating absorption edges in wide band-gap semiconductors [11]. The Ga-3*s*, 3*p* and N 1*s* absorption spectra were generated by weighting the empty pDOS with the dipole matrix-element between the core and the conduction band states. The instrumental and lifetime broadening was applied according to the same methodology used for the SXE spectra. Finally, in order to study possible anisotropy effects in the $w$-GaN phase, the theoretical SXE and SXA spectra were also computed for the N $p_z$ and N $p_{xy}$ components separately.

## 4 Results

### 4.1 N *K* x-ray emission

Figure 2 shows N 1*s* SXA-TFY spectra (top-right curves) following the N *1s->2p* dipole transitions. By measuring in both normal and grazing incidence geometries, a strong intensity variation of the unoccupied *2p$_{xy}$* and *2p$_z$* orbitals of σ* and π* symmetry are probed, respectively. The $w$-GaN crystalline film has the *c*-axis oriented parallel to the surface normal. Consequently, for SXA measured in grazing incidence geometry when the *E*-vector is almost parallel to the *c*-axis, the unoccupied out-of-plane single *2p$_z$* orbitals of π* symmetry are preferably excited and projected at the N 1*s*-edge. On the other hand, for SXA at normal incidence geometry, when the *E*-vector is orthogonal to the *c*-axis, the three tetrahedrally coordinated unoccupied 2p$_{xy}$ orbitals of σ* symmetry in the x-y plane are preferably excited and probed. The N 1*s* SXA spectra exhibit three distinguished peaks, each corresponding to the unoccupied 2p$_{xy}$ and 2p$_z$ orbitals of different weights with π* and σ* symmetry. At grazing incidence, three peak structures are observed at 400.3 eV, 402.8 eV and 405.0 eV with *2p$_z$* - π* symmetry. The measured anisotropy of the unoccupied *2p$_{xy}$* and *2p$_z$* orbitals is in good agreement with the calculated SXA spectra shown at the bottom of Fig. 2 although the peak intensities of the first two peaks at 400.3 eV and 402.8 eV are somewhat different. At normal incidence, the intensity of the central peak at 402.8 eV of σ* symmetry is much more intense than the other peaks and the 400.3 eV peak has turned into a low-energy shoulder of the main peak. The SXA spectra were used to identify at which photon energy the anisotropy is largest. The largest anisotropy is found at the main absorption peak maximum at the N 1$s$ threshold is used as excitation energy for the resonant SXE measurements (vertical tick).

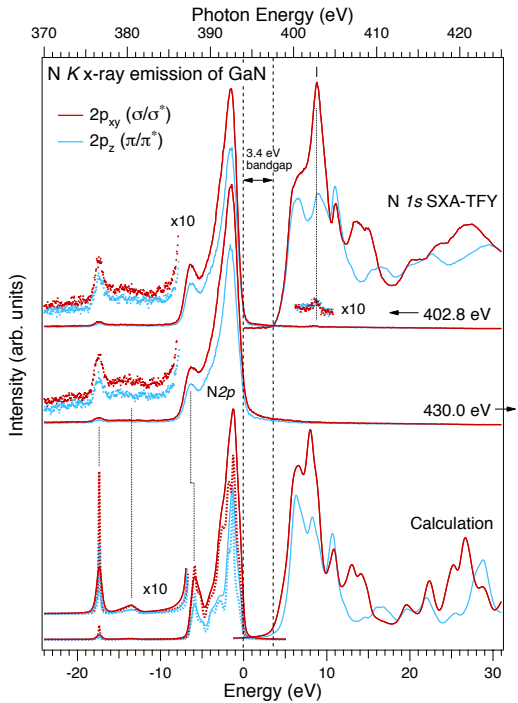

**Figure 2:** Top right: N *1s* SXA-TFY spectra and left: resonant and nonresonant N *K* SXE spectra of GaN. The SXA spectra were normalized by the step edge below and far above the N *1s* absorption thresholds. All spectra are plotted on a photon energy scale (top) and a relative energy scale (bottom) with respect to the top of the valence band. The SXE spectra were resonantly excited at the N *1s* absorption peak maximum at 402.8 eV and nonresonant far above the N *1s* absorption threshold at 430.0 eV. The excitation energy (402.8 eV) of the resonant SXE spectrum is indicated by the vertical tick at the main peak in the SXA-TFY spectra and by the very weak elastic peak enlarged by a factor of 10. To account for the experimental bandgap, the calculated SXA-TFY spectra were shifted by +1.84 eV. The calculated N *2p* charge occupation of $w$-GaN is 2.858e.

N $K$ SXE spectra following the N *2p->1s* dipole transitions of *w*-GaN, excited at 402.8 and 430.0 eV photon energies are shown in the left part of Fig. 2 measured at 20$^o$ and 70$^o$ incidence angles. The zero on the energy scale at the bottom of Fig. 2 was set by aligning the experimental and calculated N $K$ SXE spectra, whereas the relative position between the experimental SXA and SXE spectra was fixed using the elastic peak in SXE corresponding to the excitation energy at the main absorption feature (vertical tick). Calculated N $K$ SXE spectra with projected *2p$_{xy}$* and *2p$_z$* symmetries are shown at the bottom-left of Fig. 2. At 20$^o$ incidence angle, the SXE spectra of *w*-GaN probes the three occupied N *2p$_{xy}$* orbitals with σ symmetry. On the contrary, at 70$^o$ incidence, the single occupied *2p$_z$* orbital with π symmetry is probed. The main *2p$_{xy}$* (*2p$_z$*) peak with σ (π) symmetry is sharp and located at -1.5 eV relative to the top of the valence band and is due to N 2$p$ - Ga 4$p$ hybridization. A well-separated and prominent low-energy shoulder located at -6.5 eV below the top of the valence band is related to N 2$p$ - Ga 4$s$ hybridization as shown in Fig. 2. The broad structure between -13 and -15 eV is due to N 2$s$ - N 2$p$ - Ga 4$s$ hybridization. A weak low-energy peak at -17.5 eV (enhanced by a factor of 10) is due to N 2$p$ - N 2$s$ - Ga 3$d$ hybridization [8,11].

The magnitude of the Rayleigh *elastic* scattering primarily depends on the surface roughness and the experimental geometry. Note that the elastic scattering (enhanced by a factor of 10) at the resonant excitation energy (402.8 eV) is very weak which is a sign of low surface roughness [32]. The calculated SXE spectra are in excellent agreement with the experimental data although the -6.5 eV low-energy shoulder is located at -6.0 eV in the WC-GGA+U calculations. The intensity of the calculated *2p$_z$* SXE spectrum with a single occupied π orbital is lower than the *2p$_{xy}$* spectrum containing three σ orbitals with further reduced intensity between the main peak and the shoulder.

Note that in *w*-GaN, the π and σ bond distances between Ga and N are comparable, where the single N-Ga bond of π symmetry along the c-axis is slightly longer (1.9021 Å) in comparison to the three N-Ga bonds with σ symmetry (1.8902 Å). This is an indication that the in-plane σ bonding is somewhat stronger than the out-of-plane π bonding leading to anisotropy in the electronic structure. The estimated band gap between the valence and conduction band edges of SXE and SXA is 3.4 ± 0.2 eV, in good agreement with optical absorption measurements [33]. Our WC-GGA-DFT calculations of the computed direct band gap [$E_g(\Gamma-\Gamma)$] amounts to 1.83 eV, which is 1.57 eV smaller than the experimental value (3.4 eV). This general deficiency of both LDA- and GGA-DFT methods has been extensively studied in the past [34, 35], and it is now well-accepted that it can be partly overcome by introducing a specific on-site Coulomb interaction [26]. However, for semiconductors, additional contribution also comes from the electron exchange-correlation discontinuity through the band gap [36]. Due to the underestimation of the *ab initio* band gap, the calculated SXA spectra were rigidly shifted by +1.57 eV in Fig. 2.

### 4.2 Ga $M_1$ and $M_{2,3}$ x-ray emission

The top panel of Fig. 3 shows experimental Ga $M_1$ SXE spectra, probing the Ga 4$p$ states and following the *4p -> 3s* dipole transitions of GaN and pure Ga. Calculated spectra are shown by the full and dashed curves for the 4$p$ valence band. As the probability for the Ga $M_1$ transitions is about two to three orders of magnitude lower than the Ga $M_{2,3}$ emission the data collection time per spectrum is about 20 h. For Ga in GaN and pure Ga, the 4$p$ valence band consists of the Ga *4p$_{xy}$* - σ and *4p$_z$* - π orbitals, oriented in- and out-of-plane (perpendicular and parallel to the surface normal **c**). Starting with GaN, the $M_1$ spectrum (within 12 eV from the top of the valence band) exhibit two main peaks observed at -1.5 eV to -3 eV, a shoulder at -6 eV and a peak at -10 eV which is not observed in the ground state calculations. For comparison, the $M_1$ spectrum of pure Ga shown below, exhibit a broad peak at +0.5 eV and a shoulder at -3 eV relative to the top of the valence band of GaN. The calculated Ga $M_1$ spectra of GaN and pure Ga are consistent with the measured data including the shift to lower energies of the GaN valence band in comparison to pure Ga. In the calculated spectra, two small peak structures are visible around -13 eV and -17.5 eV. These peaks are due to Ga $4p$ - N $2s$ and Ga $4p$ - Ga $3d$ hybridization at the bottom of the valence band, respectively.

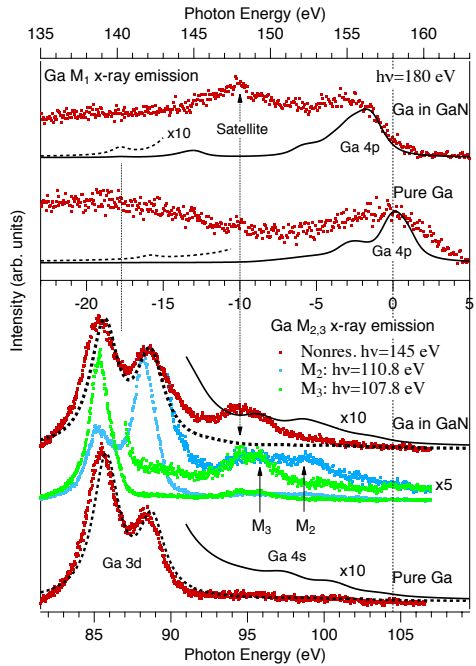

**Figure 3:** Experimental and calculated Ga $M_1$ and Ga $M_{2,3}$ SXE spectra of GaN in comparison to pure Ga. The experimental $M_1$ and $M_{2,3}$ spectra were excited nonresonantly at 180 eV and 145 eV, respectively. For GaN, resonant spectra excited at the $3p_{3/2}$ (107.8 eV) and $3p_{1/2}$ (110.8 eV) edges are also shown. A common energy scale with respect to the top of the valence band edge of GaN is indicated in the middle (vertical dotted line). The calculated Ga $4s$, $4p$ and $3d$ charge occupations of GaN: $4s$: 0.515e, $4p$: 0.474e, $3d$: 9.847e and for pure Ga $4s$: 0.693e, $4p$: 0.307e, $3d$: 9.766e. The area for the $M_2$ component was scaled down by the experimental branching ratio and added to the $M_3$ component. The calculated Ga $M_1$ spectrum of pure Ga was shifted by +1.71 eV corresponding to the calculated core-level shift between GaN and pure Ga.

The bottom panel of Fig. 3 shows experimental and calculated Ga $M_{2,3}$ SXE spectra of GaN and pure Ga probing the occupied $4s$ and $3d$ states. Calculated spectra are shown by the full and dashed curves, both for the $4s$ states (full curves) and the $3d$ spin-orbit split semicore level (dashed curves). The $4s$ states within 12 eV from the top of the valence band, consists of a double structure with the $M_3$ and $M_2$ emission of the valence band identified in the resonant spectra which are enhanced by a factor of 5. The $M_3$ and $M_2$ emission peaks (FWHM=1.4 eV) are located at -9 eV ($M_3$) and -6 eV ($M_2$), respectively. As in the case of the $M_1$ emission in the upper panel, a satellite peak structure also occurs at -10 eV in GaN for the nonresonant excitation which is absent in pure Ga. As is evident from the resonant spectrum excited at 107.8 eV, the satellite appears at -1 eV below the $M_3$ emission peak and has another origin than the normal emission as discussed in section V. The position of the Ga $4s$ states indicate hybridization between the Ga $4s$ and the N $2p$ states corresponding to the -6 eV peak in the N $K$ emission. Note the intensity between -12 to -16 eV in the $M_3$ spectrum excited at 107.8 eV which may be attributed to N $2s$ - Ga $4s$ hybridization. The spin-orbit split $3d$ level at the bottom of the valence band consists of the Ga $M_{4,5} \rightarrow M_3$ and $M_4 \rightarrow M_2$ ($3d_{5/2,3/2} \rightarrow 3p_{3/2,1/2}$) transitions with energies of -19 and -16 eV relative to the top of the valence band. Note that the $3d$ semicore level occurs at the same emission energy for both GaN and Ga with the same spin-orbit splitting of 3.1±0.1 eV. The measured Ga $M_{2,3}$ spin-orbit splitting (3.1 eV) is smaller than the calculated WC-GGA value of 3.629 eV and the calculated $3d$ core levels (dashed curves) are also closer to the E by 4.0 eV. By applying WC-GGA+U with a U-value of 10 eV, the spin-orbit splitting was optimized according to the experimental value (3.1 eV) and the $3d$ core level was shifted down by -4.0 eV. The measured $M_3/M_2$ branching ratio of Ga in GaN is 1.285:1 and for pure Ga it is 1.674 which are both smaller than the statistical single-particle ratio of 2:1. The / branching ratio is a signature of the degree of ionicity. For conducting and metallic systems like pure Ga, the / ratio is higher than the statistical ratio 2:1, but for the GaN semiconductor, the Coster-Kronig process is quenched by the band gap [37,38,39]. The larger $M_3/M_2$ ratio for pure Ga is thus a signature of better conductivity than for the GaN semiconductor.

## 5 Discussion

From the SXE spectra in figures 2 and 3 we distinguished several hybridization regions giving rise to the N 2$p$ - Ga 4$p$ and N 2$p$ - Ga 4$s$ bonding at -1.0 to -2.0 eV and -5.5 to -6.5 eV below the top of the valence band, respectively. In addition, N 2$s$ - N 2$p$ - Ga 4$s$ and N 2$s$ - N 2$p$ - Ga 3$d$ hybridization regions are found at the bottom of the valence band between -13 and -15 eV and between -17 and -18 eV, respectively. The N 2$s$ - N 2$p$ - Ga 4$s$ and N 2$s$ - N 2$p$ - Ga 3$d$ hybridization and covalent bonding are both deeper in energy from the top of the valence band than the N 2$p$ - Ga 4$p$ and N 2$p$ - Ga 4$s$ hybridization regions indicating relatively strong bonding although the intensities are lower. The peak features that we identified in the SXE spectra agree very well with those from the theoretical calculations. As observed in Fig. 4ab, the occupied states are dominated by the N 2$p$ orbitals while the unoccupied states are mainly due to the Ga 4$s$ and N 2$p$ states. Electronic dipole transitions over the band gap thus essentially involve the N 2$p$ states of the valence band and the unoccupied Ga 4$s$-states of the conduction band. In comparison to AlN [10], the Ga 3$d$ semicore level generally stabilizes the crystal structure of $w$-GaN as no 3$d$ states are located at the very top of the valence band edge in GaN. Furthermore, by looking at Fig. 4ab we address the peak feature denoted "satellite" in the Ga $M_{2,3}$ SXE spectrum of GaN (Fig. 3, bottom panel) as a feature with non-ground state property. Fig. 4a (left panel) clearly shows that there are no calculated bands in the energy region around -10 eV. We suggest that the satellite peak observed in the experimental Ga $M_{2,3}$ spectrum may be attributed to ionic Ga 4$s$ states due to charge-transfer to the N 2$p$ band. The ionic bonding in GaN thus causes a low-energy *charge-transfer satellite excitation* as previously observed in insulating systems such as NiO and CoO [40,41]. This type of satellite can be reproduced using a ligand field treatment within the Single Impurity Anderson Model (SIAM) including full multiplet configuration interaction to the Ga $4s^2L^{-1}$ state, where $L^{-1}$ denotes a hole in the N 2$p$ valence band. The energy position of charge-transfer satellites appears about -1 eV lower than the normal emission band with the characteristics of delocalized states with almost constant emission energy as the excitation energy is changed. Due to hybridization with the Ga 4*p* level, the satellite also appear as a peak feature in the Ga $M_1$ spectrum of GaN in the top panel. Note that the satellite hardly appears in the spectra of pure Ga where the ionicity and charge-transfer is much smaller or negligible.

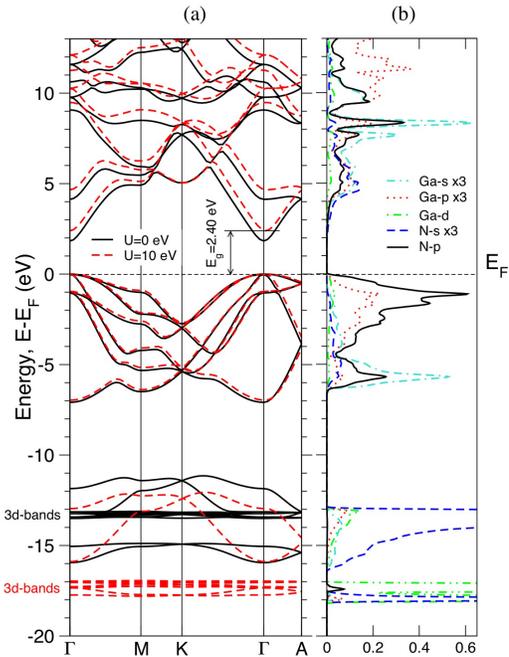

**Figure 4:** (a) Calculated ground-state band dispersions in GaN along the Γ→M→K→Γ→A direction for U=0 eV (black solid lines) and U=10 eV (red dashed lines). (b) Partial density of states as computed for the U=10 eV case by using the relaxed unit cell geometry.

Figure 4 (a) shows the differences between the calculated band-structure of GaN by using both the U=0 eV and the U=10 eV treatment on the Ga 3*d*-states. The two sets of band-dispersions were obtained at their corresponding ground state geometries. For U=10 eV, the equilibrium lattice constant and *c*/*a* ratio were completely re-determined and found to be slightly smaller (*a*=3.1565 Å, *c*/*a*=1.626) with respect to the U=0 eV case. The new set of crystal parameters lead to a theoretical bulk modulus of 206 GPa, which is about 13 GPa larger than the value computed within the U=0 eV method. As a result, the introduction of an additional potential on the Ga 3*d*-level within the WC-GGA+U formalism permits an excellent reproduction of the measured GaN SXE spectra and also improves considerably the agreement with the reported experimental elastic properties.

Our band-structure calculations with U=10 eV shows two important differences in comparison to the U=0 eV case, i) the repositioning of the Ga 3*d*-states to lower energies and ii) a band-gap opening at the Γ point by 0.57 eV. As mentioned in Sec. IIIA, we explicitly used the WC-GGA+U potential to move the position of the Ga 3*d* semicore states down to -17.2 eV, in comparison to the value of -13.2 eV given by both WC-GGA and LDA functionals [26]. The introduction of the Coulomb interaction makes the Ga 3*d* band broader (1-2 eV width) and more delocalized around the Ga atoms (see fig. 4a). This is a direct consequence of the weakening of the N 2*p* - Ga 3*d* coupling in both the WC-GGA+U and the LDA+U [26] approaches. Although it does not fully account for the energy-gap discrepancy, the WC-GGA+U method also opens up the theoretical $E_g$ to 2.40 eV, thus slightly improving the agreement with the experimental band-gap.

The WC-GGA (LDA) $E_g$ shortcoming can thus mainly be attributed to an incorrect positioning of Ga 3*d*-levels, which are not deep enough in energy and therefore cannot be considered chemically inert. As a consequence, the 3*d* levels over-interact with the valence-band states implying an energy gap closing [42,43]. Not even quasiparticle calculations within the GW approach [36] involving essentially long-range dynamic correlations can fully correct for the low binding energy of the shallow 3*d* semicore level. For other wide band gap semiconductors such as AlN [10], the *p* - *d* interaction plays a smaller role due to the fact that the 3*d* band is only partially occupied and the main reason for the underestimation of the band gap therefore occurs from a discontinuity in the electron exchange-correlation potential [36]. For more exact addressing to what extent the Coulomb interaction affects the valence-band maximum and the conduction-band minimum, the same calculation procedure used for band lineups in semiconductor heterojunctions needs to be performed [44]. Recently, a few studies made use of this methodology and tried to address the DFT+U band gap opening in GaN via either a modified screening effect, that essentially moves the conduction band minimum [26], or through a mere shifting of the valence band alone [29]. The position of the valence band maximum and the influence of the screening effect can be systematically studied by substitution and alloying of Ga in GaN to Al in AlN that withdraws charge in the *sp* valence band. This kind of studies needs further work and may be addressed in the future. As in the case of Ge [45], the shallow Ga 3*d* semicore level at -17.5 eV interacts with the 4*sp* valence band and withdraws charge density that opens the band gap.

## 6 Conclusions

The electronic structure and chemical bonding of (0001)-oriented single crystal wurtzite GaN in comparison to pure Ga has been investigated. The combination of element selective soft x-ray absorption, soft x-ray emission spectroscopy and electronic structure calculations show that in GaN, the main N 2*p* - Ga 4*p* and N 2*p* - Ga 4*s* hybridization and bond regions appear -1.0 to -2.0 eV and -5.5 to -6.5 eV below the top of the valence band, respectively. In addition, N 2*p* - N 2*s* - Ga 4*s* and N 2*s* - N 2*p* - Ga 3*d* hybridization regions are found at the bottom of the valence band between -13 and -15 eV, and -17 to -18 eV. The Ga $M_{2,3}$ emission spectra of GaN as compared to pure Ga exhibit a band-like charge-transfer satellite feature at -10 eV which does not exist in the calculated ground state spectra. A strong polarization dependence in both the occupied and unoccupied N 2*p* states is identified as due to different weights of the *2p$_{xy}$* and *2p$_z$* orbitals of σ/σ* and π/π* symmetry, respectively. The major materials property effect of the change of the electronic structure, charge occupation and the chemical bond regions in comparison to pure Ga is the opening of the wide band gap of 3.4 eV in GaN. Including on-site Coulomb correlation interactions for the 3*d* semicore states using the WC-GGA+U functional with U=10 eV we were able to improve the description of the spectral peak positions and the hybridization regions of GaN and thus determine a widening of the GaN band gap at the Γ point. Moreover, by using a finite U-value we also compute the bulk modulus of GaN to be in better agreement with the experimental hardness. Our calculations show that the Coulomb interaction lowers the energy of the 3*d* states by about -4 eV implying a certain weakening of the N 2*p* - Ga 3*d* coupling. It also makes the width of the semicore 3*d* states broader as the 3*d* electrons become more delocalized around the Ga atoms. These two effects can explain the energy gap opening in Ga-based semiconductors.

## 7 Acknowledgements

We would like to thank the staff at MAX-lab for experimental support and Anders Lundskog for providing the GaN sample. This work was supported by the Swedish Research Council Linnaeus Grant LiLi-NFM, the Göran Gustafsson Foundation, the Swedish Strategic Research Foundation (SSF), Strategic Materials Reseach Center on Materials Science for Nanoscale Surface Engineering (MS$^2$E). One of the authors (M. Mattesini) wishes to acknowledge the Spanish Ministry of Science and Technology (MCyT) for financial support through the *Ramón y Cajal* program.

## References

[1] *Diamond, Silicon Carbide and Nitride Wide Bandgap Semiconductors*, Mater. Res. Soc. Symp. Proc., edited by C. H. Carter, Jr. (Materials Research Society, Pittsburgh, PA, 1994), Vol. 339.

[2] M. Misra, T. D. Moustakes, R. P. Vaudo, R. Sigh and K. S. Shan, in *III-V Nitrides*, edited by F. A. Ponce, T. D. Moustakas, I. Akasaki and B. A. Monemar (materials Research Society, Pittsburgh, 1996), Vol. 449, p. 597.

[3] K. E. Smith and S. D. Kevan, Prog. Solid State Chem. **21**, 49 (1991).

[4] S. A. Ding, G. Neuhold, J. H. Weaver, P. Höberle, K. Horn, O. Brandt, H. Yang and K. Ploog; J. Vac. Sci. Technol. A **14**, 819 (1996).

[5] W. R. L. Lambrecht, B. Segall, S. Strite, G. Martin, A. Agarwal, H. Morkoc and A. Rockett; Phys. Rev. B **50**, 14155 (1994).

[6] K. Lawniczak-Jablonska, T. Suski, Z. Liliental-Weber, E. M. Gullikson, J. H. Underwood, R. C. C. Perera and T. J. Drummond, Appl. Phys. Lett. **70**, 2711 (1997).

[7] K. E. Smith, L.-C. Duda, C. B. Stagarescu, J. Downes, D. Korakakis, R. Singh, T. D. Moustakas, J. Guo and J. Nordgren; J. Vac. Sci. Technol. B **16**, 2250 (1998).

[8] L.-C. Duda, C. B. Stagarescu, J. Downes, K.-E. Smith, D. Korakakis, T. D. Moustakas, J. Guo and J. Nordgren; Phys. Rev B **58**, 1928 (1998).

[9] C. B. Stagarescu, L.-C. Duda, K. E. Smith, J. H. Guo, J. Nordgren, R. Singh and T. D. Moustakas; Phys. Rev B **54**, R17335 (1996).

[10] M. Magnuson, M. Mattesini, C. Höglund, J. Birch and L. Hultman; Phys. Rev B **40**, 155105 (2009).

[11] V. N. Strocov, T. Schmitt, J.-E. Rubensson, P. Blaha, T. Paskova, and P. O. Nilsson, Phys. Rev. B **72**, 085221 (2005).
[12] A. Wu and R. Cohen, Phys. Rev. B **73**, 235116 (2006).
[13] F. Tran, R. Laskowski, P. Blaha and K. Schwarz, Phys. Rev. B **75**, 115131 (2007).
[14] A. Kakanakova-Georgieva, U. Forsberg, I. G. Ivanov and E. Janzen; J. Crystal Growth; **300**, 100 (2007).
[15] U. Forsberg, A. Lundskog, A. Kakanakova-Georgieva, R. Ciechonski and E. Janzen; J. Crystal Growth; **311**, 3007 (2009).
[16] P. Fewster and N. Andrews, J. Appl. Cryst. **28**, 451 (1995).
[17] H. E. Swanson, R. K. Fuyat, Natl. Bur. Stand. (U.S.) Circular 539, II, 9 (1953) (Data base ref. 05-0601).
[18] R. Denecke, P. Vaterlein, M. Bassler, N. Wassdahl, S. Butorin, A. Nilsson, J.-E. Rubensson, J. Nordgren, N. Mårtensson and R. Nyholm; J. Electron Spectrosc. Relat. Phenom. **101-103**, 971 (1999).
[19] J. Nordgren and R. Nyholm; Nucl. Instr. Methods **A246**, 242 (1986); J. Nordgren, G. Bray, S. Cramm, R. Nyholm, J.-E. Rubensson and N. Wassdahl; Rev. Sci. Instrum. **60**, 1690(1989).
[20] P. Blaha, K. Schwarz, G. K. H. Madsen, D. Kvasnicka and J. Luitz, WIEN2k, An Augmented Plane Wave + Local Orbitals Program for Calculating Crystal Properties (Karlheinz Schwarz, Techn. Universität Wien, Austria), 2001. ISBN 3-9501031-1-2.
[21] P. Hohenberg and W. Kohn, Phys. Rev. **136**, B864 (1964).
[22] W. Kohn and L. J. Sham, Phys. Rev. **140**, A1133 (1965).
[23] E. Sjöstedt, L. Nordström, and D. J. Singh, Solid State Commun. **114**, 15 (2000).
[24] *Semiconductors-Basic Data*, 2nd revised ed., edited by O. Madelung (Springler, Berlin, 1996).
[25] H. Schulz and K. H. Thiemann, Solid State Commun. **23**, 815, (1977).
[26] A. Janotti, D. Segev, and C. G. Van de Walle, Phys. Rev. B **74**, 045202 (2006).
[27] F. Birch, Phys. Rev. **71**, 809, (1947).
[28] M. D. Drory, J. W. Ager III, T. Suski, I. Grzegory, and S. Porowski, Appl. Phys. Lett., **69**, 4044 (1996).
[29] R. Cherian, P. Mahadevian and C. Persson; Soldi State Comm **149**, 1810 (2009).
[30] U. von Barth, and G. Grossmann, Phys. Rev. B **25**, 5150 (1982).
[31] D. A. Muller, D. J. Singh, and J. Silcox, Phys. Rev. B **57**, 8181 (1998).
[32] M. Magnuson, S. M. Butorin, C. Såthe, J. Nordgren and P. Ravindran; Europhys. Lett. **68**, 289 (2004).
[33] R. Laskowski and N. E. Christensen, Phys. Rev. B **74**, 075203 (2006).
[34] J. P. Perdew, R. G. Parr, M. Levy, and J. L. Balduz, Phys. Rev. Lett. **49**, 1691 (1982).
[35] J. P. Perdew, M. Levy, and J. L. Balduz, Phys. Rev. Lett. **51**, 1884 (1983).
[36] W. Ku and A. G. Eguiluz; Phys. Rev. Lett. **89**, 126401 (2002).
[37] M. Magnuson, N. Wassdahl and J. Nordgren; Phys. Rev. B **56**, 12238 (1997).
[38] M. Magnuson, L.-C. Duda, S. M. Butorin, P. Kuiper and J. Nordgren; Phys. Rev. B **74**, 172409 (2006).
[39] M. Magnuson, N. Wassdahl, A. Nilsson, A. Föhlisch, J. Nordgren and N. Mårtensson; Phys. Rev. B **58**, 3677 (1998).
[40] M. Magnuson, S. M. Butorin, A. Agui and J. Nordgren; J. Phys. Cond. Mat. **14**, 3669 (2002).
[41] M. Magnuson, S. M. Butorin, J.-H. Guo and J. Nordgren; Phys. Rev. B **65**, 205106 (2002).
[42] S.H. Wei and A. Zunger, Phys. Rev. B. **37** 8958 (1998).
[43] P. Schröer, P. Krüge and J. Pollmann, Phys. Rev. B. **48** 18264 (1993).
[44] C. G. Van de Walle and R. M. Martin, Phys. Rev. B **35**, 8154 (1987).
[45] M. Magnuson, O. Wilhelmsson, M. Mattesini, S. Li, R. Ahuja, O. Eriksson, H. Högberg, L. Hultman and U. Jansson; Phys. Rev. B **78**, 035117 (2008).